\documentclass[reprint,amsmath,amssymb,aps,prb,floatfix]{revtex4-2}%[reprint,amsmath,amssymb,aps,prb]{revtex4-2}
\usepackage[utf8]{inputenc} % safe with modern TeX; ok to remove if not needed
\usepackage{graphicx}
\usepackage{dcolumn}
\usepackage{bm}
\usepackage{comment}
\usepackage{amsmath}
\usepackage[safe]{tipa}
\usepackage[caption=false]{subfig}
\usepackage[hidelinks]{hyperref}
\usepackage{nameref}
\usepackage{orcidlink}
\begin{document}
	
	\preprint{APS/123-QED}
	
	\title{Sharp transitions in the spectra of {\color{black} small Frenkel-like excitons} for multi-orbital lattice systems}
	
	\author{Man-Yat Chu\orcidlink{0009-0003-7986-8709}}
	\email{mchu03@phas.ubc.ca}
	\affiliation{Quantum Matter Institute, University of British Columbia, Vancouver, British Columbia V6T 1Z4, Canada}
	
	\affiliation{Department of Physics and Astronomy, University of British Columbia, Vancouver, British Columbia, Canada,V6T 1Z1}
	
	\author{Mona Berciu}
\affiliation{Quantum Matter Institute, University of British Columbia, Vancouver, British Columbia V6T 1Z4, Canada}

\affiliation{Department of Physics and Astronomy, University of British Columbia, Vancouver, British Columbia, Canada,V6T 1Z1}
	
	\date{\today}
	
	\begin{abstract}
		\textcolor{black}{We propose a method for calculating exciton spectra and wavefunctions for model lattice Hamiltonians, based on real-space electron-hole propagators. We verify that our results agree with those of the continuum approximation in the limit of large Wannier excitons, and propose a simple criterion to estimate the exciton size above which the continuum approximation is quantitatively accurate. We then investigate simple one- and two-dimensional multi-orbital lattice models and show that small, Frenkel-like excitons, whose size approaches the lattice constant, can display physics that disagrees with the simplest continuum descriptions (a single-valley quadratic expansion around the minimum gap) not just quantitatively, but qualitatively. Specifically, we identify sharp transitions in the character and momentum of the lowest-energy exciton, enabled by the multi-orbital nature of the lattice models.}
	\end{abstract}
	
	\maketitle
	
	\section{Introduction}
	\label{sec:Intro}
	
	Excitons -- electron-hole pairs bound by their mutual Coulomb attraction---are observed in the optical response of most semiconductors. In conventional inorganic semiconductors such as Si, the strength of the attraction is small compared to the bandwidths of the conduction and valence bands, and the resulting Wannier-Mott excitons \cite{Wannier1937,Knox1963,YuCardona,OnidaReiningRubio2002} \color{black} have a radius that greatly exceeds the lattice constant $a$. Consequently, a continuum description suffices: only the states at the top of the valence band  and at the bottom of the conduction band contribute significantly to the exciton wavefunction, and their dispersion is well approximated as quadratic. \color{black} For three-dimensional (3D) crystals and for a $1/r$ Coulomb attraction, this maps the exciton spectrum to that of a hydrogen atom, with a rescaled Bohr radius and Rydberg energy \cite{Knox1963,YuCardona,OnidaReiningRubio2002,Elliott1957,RohlfingLouie2000}.

    For a strongly screened short-range attraction with a characteristic scale $U$, excitons appear in 3D only if $U$ exceeds a critical value. However, in both two dimensions (2D) and one dimension (1D), a bound state exists for arbitrarily small $U$, with a binding energy $E_{\text{bind}}\propto\exp(-W/U)$ and $E_{\text{bind}}\propto U^2/W$, respectively \cite{Elliott1957,LandauLifshitz,Kornilovitch2024} where $W$ is a measure of the average kinetic energy of the pair.  These asymptotic estimates are widely used in standard discussions of excitons \cite{HaugKoch2004,KiraKoch2011} and provide a baseline for comparison between theory and experiments, for example, see Refs.~\onlinecite{Reutzel2024, boschini2024}.
	
	However, many low-dimensional semiconductors that are technologically relevant host  small, Frenkel and charge-transfer excitons \cite{Frenkel1931,Davydov1971,Agranovich2009}  whose radii are equal or comparable to the lattice constant $a$.  For example, TR-ARPES on epitaxial C$_{60}$ films on Au(111) has revealed binding energies of 300--1000 meV  for excitons that are either on the same C$_{60}$ molecule or have the electron and hole on neighboring molecules \cite{Latzke2019,Emmerich2020,tully2024,tully2023,greenwood2025}. These results agree with previous work on C$_{60}$ using other optical techniques \cite{Lof1992,Wang1995}. {\color{black} Other relevant examples include organic crystals such as pentacene and PTCDA \cite{Knupfer2003, Bulovic1996}, and  magnetic insulators such as monolayer CrX$_3$ \cite{Wu2022, Smiertka2026, Kang2020, Dirnberger2022}.} 
    
    For such small, Frenkel-like excitons, the specific lattice symmetry becomes relevant beyond just the renormalization of the conduction and valence band effective masses \cite{Gunlycke2016}. This is because strong electron-hole attraction mixes \color{black} hole  (electron) states from across the entire valence (conduction) bands into the exciton wavefunction, so the details of the entire bands are now relevant. Furthermore, especially in organic semiconductors, the conduction and valence bands often have multi-orbital character, leading to  spectral fingerprints and momentum-dependent symmetry changes \cite{Iskin2021,Bieniek2022,Bennecke2024}. 
    
   This is why the study of small, Frenkel-like excitons requires (i) a lattice Hamiltonian that includes the relevant details of the valence and conduction bands, as well as a good description of the electron-hole attraction; and (ii) efficient methods to find the eigenenergies and eigenfunctions of eigenstates containing one particle-hole pair for this lattice Hamiltonian. 

   In this paper, we focus on aspect (ii), and propose a method adapted from studies of few-electron interacting systems \cite{MB11} to calculate the real-space electron-hole propagators, from which the exciton spectrum and wavefunction can then be extracted. Although this method can be used for excitons of any size, its efficiency improves with decreasing exciton size, so it is most suitable for the study of small, Frenkel-like excitons. As further discussed below, we believe that in this limit, it offers a  useful alternative to other methods for calculating exciton spectra, which are carried out in momentum space. 

   We exemplify this method by calculating exciton spectra for several model Hamiltonians in 1D and 2D, where the conduction and valence bands are described in terms of simple single- or multi-orbital tight-binding models, and the Coulomb attraction is assumed to be short-range. We note that these approximations are made for convenience, and we also discuss how they can be relaxed for more complex models. In terms of physics, we use these simple model Hamiltonians to investigate when and how the continuum approximation fails as the exciton's size approaches the lattice constant, and to highlight possible non-trivial  behavior of  small, Frenkel-like exciton spectra in systems with  multi-orbital bands. 

In principle, this method can also be used with realistic lattice Hamiltonians to calculate exciton spectra of specific materials, provided that accurate tight-binding models can be obtained from {\em ab-initio} methods and that the Coulomb attraction can be parameterized in a reasonably simple form, such as the Keldysh form \cite{Keldysh1979, CudazzoTokatlyRubio2011}. However, given the significant  efforts and progress in developing efficient schemes for {\em ab-initio} simulations of excitons and their optical response, it is not clear to us whether our method would out-perform other  approaches to solving the electron-hole Bethe-Salpeter equation, implemented in existing {\em ab-initio} codes \cite{OnidaReiningRubio2002,CudazzoTokatlyRubio2011, BerkelbachPRB2013, ChernikovPRL2014, WuPRB2015, GalvaniPRB2016,CudazzoPRL2016}. The comparison might become favorable in the limit of small, Frenkel-like excitons, provided that the extraction of the parameters of the real-space Hamiltonian needed as the input for our method is not too computationally expensive. 

In practice, we believe that the method we propose here is most suitable to study properties of small, Frenkel-like excitons in model Hamiltonians in order to gain intuition about the influence of the various terms of the Hamiltonian on their spectra and possible behaviors. These model Hamiltonians are an intermediary level of modeling between the simple continuum approximation (which works very well for Wannier-like excitons, but fails for small, Frenkel-like excitons) and full {\em ab-initio} simulations of excitons in specific materials.

Moreover, because our method is formulated in real space, one can generalize it  beyond model Hamiltonians that are translationally invariant, to also study behavior of excitons near defects and/or interfaces which break translation symmetry. Another possible direction is to generalize the model Hamiltonian by adding Holstein or other types of model carrier-phonon couplings, to investigate their generic effects on the resulting exciton-polaron, as done in Ref. \onlinecite{Stepan} for an extremely simple model.

\color{black}

%    These considerations show that the convenient, simple estimates provided by the continuum approximation must become quantitatively inaccurate once {\color{purple} small Frenkel-like exciton}s are concerned, yet it is generally believed that they remain qualitatively correct. In this work, we show that it is very easy to find examples where these simple estimates are {\it qualitatively} wrong, especially for {\color{purple} small Frenkel-like exciton}s in multi-orbital models. Fortunately, solving such two-particle problems (here, the electron and hole) on an infinite lattice can be done very efficiently, as we demonstrate with a method adapted from studies of few-particle interacting systems, see {\it e.g.} Ref. \onlinecite{MB11}. We hope that such methods will be adopted to study realistic lattice models and obtain accurate properties of their {\color{purple} small Frenkel-like exciton}s.

    The article is organized as follows: Section \ref{sec:lat_model} introduces the lattice model Hamiltonians we study, and Section \ref{sec:formalism} details our formalism. Our results are shown in Section \ref{sec:Results}, and we summarize our findings in Section \ref{sec:conclusion}.

	\section{The lattice models}
    \label{sec:lat_model}

The generic Hamiltonian describing an exciton is:
\begin{equation}
{\cal H} = {\cal H}_e +  {\cal H}_h + V_{\rm e-h}
\label{1}
\end{equation}
where the terms describe the electron moving in the otherwise empty conduction band, the hole moving in the otherwise full valence band, and their Coulomb attraction, respectively. 

In our lattice models, we allow the valence band to be multi-orbital:
\begin{equation}
{\cal H}_h =  -\sum_{\langle n, m \rangle, \alpha \beta} t_{\alpha \beta} ({\bf R}_n-{\bf R_m}) (v^\dagger_{\alpha n} v_{\beta m} + H.c.) 
\label{2}
\end{equation}
Here $n,m$ index lattice sites located at $ {\bf R}_n, {\bf R_m}$, $\alpha, \beta$ label the specific valence orbitals, and $v^\dagger_{\alpha n}$ creates an electron in the orbital $\alpha$ at site $n$. Explicitly, we will show results for 1D models with 2 orbitals per site ($1s$ and $2p_x$ for a chain parallel to the $x$ axis) and 2D triangular models with 3 orbitals per site ($d_{x^2-y^2}, d_{xy}, d_{3z^2-r^2}$ for a lattice lying in the $xy$ plane). Any other combinations of orbitals can be studied similarly. Crystal fields can be included trivially, but we ignore them for simplicity. We restrict the hopping to nearest-neighbor for simplicity, but this can be generalized straightforwardly to longer range. To minimize the number of free parameters for the 2D system, we use the Harrison rules \cite{Harrison1989} to calculate ratios between the hopping integrals, see Appendix \ref{App:hopping} for details. For the same reasons, we present results only for a single-orbital conduction band:
\begin{equation}
{\cal H}_e = - t_c \sum_{\langle n, m \rangle} (c^\dagger_n c_m + H.c.) + \Delta \sum_n c^\dagger_n c_n 
\label{3}
\end{equation}
where $c_n^\dagger$ creates a conduction-band electron at site $n$ of the lattice. Here $\Delta$ is the difference between the on-site energies and controls the value of the band gap. For simplicity we consider again only nearest-neighbor hopping; generalization to longer range hopping is straightforward. \color{black} Similarly, generalization to a multi-orbital conduction band is straightforward;  this introduces more free parameters, but it can be dealt with similarly to how we treat the multi-orbital valence band.  

We emphasize that here we ignore possible hybridization between the $c_n$ and $v_{\alpha m}$ orbitals, which can be a reasonable approximation for a large gap semiconductor under the right circumstances. Including such hybridization complicates matters because the real-space counterparts of the conduction and valence band Bloch eigenstates are no longer the atomic orbitals $c_n$ and $v_{\alpha m}$, but instead more extended Wannier orbitals. While dealing with such complications is beyond the scope of this work, we believe that they can be handled with the general method described below, and among other things, will give rise to an exchange interaction that will lift the degeneracy between singlet and triplet excitons which holds for the simpler model Hamiltonians discussed here. \cite{CudazzoPRL2016,Qiu2015} In the absence of such hybridization, the spin degrees of freedom are trivial, and from now on we drop the spin labels in this work. 

\color{black}

Hamiltonians (\ref{2}) and (\ref{3})  can be diagonalized straightforwardly to:
\begin{equation}
{\cal H}_h =  \sum_{{\bf k}, \gamma} E_{v}({\bf k}, \gamma)v^\dagger_{{\bf k}, \gamma} v_{{\bf k}, \gamma}
\label{4}
\end{equation}
where ${\bf k}$ is the momentum restricted to the Brillouin zone and $\gamma$ indexes the valence bands of energy $E_{v}({\bf k},\gamma)$, and
\begin{equation}
{\cal H}_e =  \sum_{{\bf k}} E_{c}({\bf k})c^\dagger_{{\bf k}} c_{{\bf k}}
\label{5}
\end{equation}
defines the conduction band energy $E_{c}({\bf k})$.

We describe the electron-hole attraction as:
\begin{equation}
V_{\rm e-h} = -\sum_{n,\alpha} U_\alpha c^\dagger_n c_n v_{\alpha,n} v^\dagger_{\alpha,n} 
\label{6}
\end{equation}
At first sight, this on-site attraction is an unreasonable approximation for organic semiconductors whose {\color{black} small, Frenkel-like exciton}s are precisely due to the very weakly screened long-range Coulomb interaction. Nevertheless, in the limit of on-site Frenkel excitons of most interest to us in this work, the electron and hole are on the same site with extremely high probability. Because they never 'explore' longer separations, the details of the longer-range Coulomb attraction become irrelevant. The only meaningful interaction energy scale in this limit is the difference between the attraction between an on-site pair and one occupying adjacent sites. This is what the $U_\alpha$ energies describe. For completeness, we mention that longer range Coulomb attraction can also be studied with the method we discuss below. A sample calculation for the effect of adding a nearest neighbor interaction can be found in Appendix~\ref{App:OtherU}.

\section{The method}
\label{sec:formalism}

Because of the translational invariance of the Hamiltonian (\ref{1}), the exciton total momentum ${\bf K}$ is a good quantum number; its value is restricted to the Brillouin zone of the appropriate lattice. For any given value of ${\bf K}$, a particularly convenient basis for calculating {\color{black} small, Frenkel-like exciton} eigenstates is the set of all states \cite{Knox1963,HaugKoch2004,MB11}: 
\begin{equation}
|\alpha, {\bf K}, \bm{\delta} \rangle = \sum_n {e^{i{\bf K}({\bf R}_n+\bm{\delta}/2)} \over \sqrt{N}} c^\dagger_n v_{\alpha, n+\bm{\delta}} |{\rm GS}\rangle
\label{7}
\end{equation}
where $|{\rm GS}\rangle$ is the ground state with filled valence bands and empty conduction band, and $N\rightarrow \infty$ is the number of lattice sites. \color{black} This is because for {\color{black} small, Frenkel-like exciton}s, the relative displacement $\bm{\delta}$ between the electron and hole is short, meaning that  the eigenfunction of an exciton with momentum ${\bf K}$:
\begin{equation}
|{\bf K}, {\rm exc} \rangle = \sum_{\alpha, \bm{\delta}} \phi_{\bf K}(\alpha, \bm{\delta} )|\alpha, {\bf K}, \bm{\delta} \rangle 
\label{7b}
\end{equation}
will have significant amplitude $\phi_{\bf K}(\alpha, \bm{\delta} )$ only for a few values of $\bm{\delta}$, like  $\bm{\delta}=0$ when the electron and hole are at the same site, $|\bm{\delta}|=a$ where the electron and hole are on neighboring sites, etc. Because the electron and hole are bound, there is certainly a cutoff value $\delta_M$ beyond which the probability $|\phi_{\bf K}(\alpha, \bm{\delta} )|^2$ to find the electron and hole at a displacement $|\bm{\delta}|\ge  \delta_M$ becomes so small that it can be set to zero without loss of accuracy. The smaller the exciton, the smaller the value of $\delta_M$, and thus fewer coefficients $\phi_{\bf K}(\alpha, \bm{\delta} )$ with  $|\bm{\delta}|\le \delta_M$ are needed to reconstruct an accurate exciton eigenfunction.

This should be contrasted with the more established approach of working entirely in the momentum space. In this case, 
the eigenfunction for an exciton with momentum ${\bf K}$ can be rewritten as:
\begin{equation}
|{\bf K}, {\rm exc} \rangle = \sum_{\alpha, {\bf k}} \phi_{\bf K}(\alpha, {\bf k}-{{\bf K}\over 2} )c^\dagger_{\bf k}
v_{\alpha, \bf{k} - \bf{K}} |{\rm GS}\rangle
\label{7d}
\end{equation}
where $\phi_{\bf K}(\alpha, {\bf k}-{{\bf K}\over 2} )= \sum_{\bm{\delta}} e^{i ({\bf k}-{{\bf K}\over 2} )\bm{\delta}}/\sqrt{N} \phi_{\bf K}(\alpha, \bm{\delta} )$ is the Fourier transform of the real-space relative wavefunction  $\phi_{\bf K}(\alpha, \bm{\delta} )$. The more localized is the real space wavefunction $\phi_{\bf K}(\alpha, \bm{\delta} )$, the more extended over the full Brillouin zone becomes $\phi_{\bf K}(\alpha, {\bf k}-{{\bf K}\over 2} )$. In other words, to reconstruct an accurate  small, Frenkel-like exciton eigenstate, one needs to calculate accurate values for $\phi_{\bf K}(\alpha, {\bf k}-{{\bf K}\over 2} )$ in the entire Brillouin zone. Reaching convergence requires defining an appropriate grid of ${\bf k}$ points in the Brillouin zone; accuracy improves with an increasing number of grid-points, but this also increases computational costs.   This is why for sufficiently  small excitons, the real-space basis of Eq. \ref{7} eventually becomes more efficient.

Conversely, as an exciton approaches the Wannier limit, working in the real-space basis becomes inefficient because $\phi_{\bf K}(\alpha, \bm{\delta} )$ now spreads over many sites. In this limit, working in the momentum space becomes favorable because here $\phi_{\bf K}(\alpha, {\bf k}-{{\bf K}\over 2} )$ is large only for momenta ${\bf k}$ near the bottom of conduction band.  

We emphasize that the two approaches are  fully equivalent and both will produce the same exciton spectra, the only difference being the computational cost to obtain accurate results. Next, we briefly explain how we find the exciton spectrum using the real-space basis.

\color{black}

Instead of solving the Schr\"odinger equation directly, it is more convenient to calculate instead the particle-hole real-space propagators \cite{MB11}:
\begin{equation}
G_{\alpha,\beta}({\bf K}, \bm{\delta}, z) = \langle \alpha, {\bf K}, \bm{0} | \hat{G}(z) | \beta, {\bf K}, \bm{\delta} \rangle
\label{8}
\end{equation}
where \(\hat{G}(z) = (z - H)^{-1}\) is the resolvent, $z=\omega+i\eta$ is the energy plus a broadening $\eta\rightarrow 0$ and we set $\hbar=1$  and $\eta = 10^{-2}t_c$ throughout unless otherwise specified. These propagators are useful because according to the Lehmann representation \cite{Lehman1954,Mahan2000}:
\begin{equation}
G_{\alpha,\beta}({\bf K}, \bm{\delta}, z) = \sum_{\lambda} \frac{\langle \alpha, {\bf K}, \bm{0} | {\bf K},\lambda\rangle\langle  {\bf K},\lambda | \beta, {\bf K}, \bm{\delta} \rangle}{z- E({\bf K},\lambda)}
\label{9}
\end{equation}
where ${\cal H}|{\bf K},\lambda\rangle= E({\bf K},\lambda)|{\bf K},\lambda\rangle$ are the eigenstates of ${\cal H}$ with a total momentum ${\bf K}$ and which contain one electron-hole excitation, with $\lambda$ labeling the remaining quantum numbers.  

In the absence of Coulomb interactions, these electron-hole eigenstates are $ c^\dagger_{\bf k}v_{{\bf k-K}, \gamma}|{\rm GS}\rangle$ for any ${\bf k}$, with the corresponding energy $E_c({\bf k})-E_v({\bf k-K},\gamma)$. They define the electron-hole continuum that for a given ${\bf K}$ starts at 
\begin{equation}
E_{\rm gap}({\bf K})= \min_{{\bf k}, \gamma} [E_c({\bf k})-E_v({\bf k-K},\gamma)]
\label{10}
\end{equation}
The semiconductor gap is then $E_{gap}= \min_{\bf K} E_{\rm gap}({\bf K})$, and it is a direct gap if the minimum is reached at ${\bf K}=0$; otherwise, it is an indirect gap.

If the Coulomb attraction is sufficiently strong to bind excitons (which is always the case in 1D and 2D systems), their eigenenergies are signaled by discrete poles of $G_{\alpha,\beta}({\bf K}, \bm{\delta}, z) $ at energies inside the gap $\omega = E_{\rm exc}({\bf K}) < E_{\rm gap}({\bf K})$. \textcolor{black}{The presence of such discrete states signals stable bound electron-hole (excitons) eigenstates below the particle-hole continuum}.
The corresponding exciton binding energy is $E_{\rm bind}({\bf K}) =  E_{\rm gap}({\bf K})- E_{\rm exc}({\bf K})$ \cite{OnidaReiningRubio2002,Mahan2000}.

Furthermore, the residues associated with each such pole give direct access to that exciton's wavefunction in real space, because
\begin{equation}
G_{\alpha,\beta}({\bf K}, \bm{\delta}, z)\big|_{\omega \rightarrow E_{\rm exc}({\bf K})} \approx \frac{\langle \alpha, {\bf K}, \bm{0} | {\bf K}, {\rm exc}\rangle\langle  {\bf K},{\rm exc} | \beta, {\bf K}, \bm{\delta} \rangle}{z - E_{\rm exc}({\bf K})}
\label{11}
\end{equation}
meaning that (see Eq. (\ref{7b})):
\begin{equation}
\phi^*_{\bf K}(\beta, \bm{\delta})=\langle  {\bf K},{\rm exc} | \beta, {\bf K}, \bm{\delta} \rangle \propto G_{\alpha,\beta}({\bf K}, \bm{\delta}, E_{\rm exc}({\bf K})). 
\label{12}
\end{equation}
Eq. (\ref{12}) gives the real-space exciton amplitude  after normalization $\sum_{\beta,\bm \delta}| \phi_{\bf K}(\beta, \bm{\delta})|^2 =1$. We also use it to extract the exciton radius $\xi $ from the long-range envelope $  \phi_{\bf K}(\beta, \bm{\delta})\rightarrow\exp(-|\bm{\delta}|/\xi)$ when $|\bm{\delta}| \rightarrow \infty$ \cite{Economou2006}.

These real-space particle-hole propagators are calculated from matrix elements between states of Eq.~ (\ref{7}) of the identity $\hat{G}(z)(z-{\cal H}) =1$, from which we find:
\begin{equation}
z G_{\alpha,\beta}({\bf K}, \bm{\delta}, z) = \delta_{\alpha \beta} \delta_{\bm{\delta},\bm{0}} + \langle \alpha, {\bf K}, \bm{0} | \hat{G}(z) {\cal H}| \beta, {\bf K}, \bm{\delta} \rangle
\label{13}
\end{equation}
The second term can be expanded in terms of other real-space propagators, \color{black} in particular for models with only nearest-neighbor hopping considered here, the right hand side will also contain propagators where the electron-hole displacement is either increased or decreased by one lattice constant, due to hopping.   We can then generate a linear system of coupled equations linking all these propagators, from which we can calculate their values for any energy. In principle, the size of this system is infinite because $\bm{\delta}$ can be arbitrarily large, but as already discussed, for small, Frenkel-like excitons the magnitude of the propagators corresponding to energies inside the gap decreases very fast with increasing $\bm{\delta}$  \cite{Economou2006}. This allows us to truncate this system by setting all propagators to zero with $|\bm{\delta}|\ge  \delta_M$, where the cutoff $\delta_M$  is increased until the exciton energy is converges within the desired accuracy. For very  small, Frenkel  excitons, as shown below, reasonable accuracy may be obtained even for a cutoff  $\delta_M=2a$, making such calculations extremely efficient. More details and the explicit equations for these propagators are given in Appendix \ref{sec:Conv_1D}.

\color{black}

\begin{figure}[t]
\centering
\includegraphics[width=0.485\linewidth]{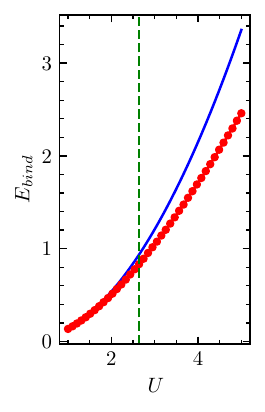} 
\includegraphics[width=0.49\linewidth]{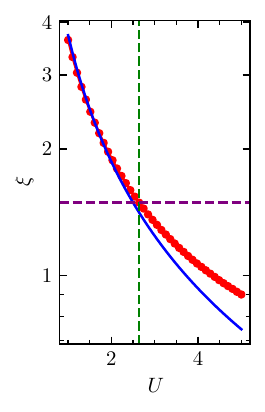} 
\caption{\textcolor{black}{Binding energy of the 1D exciton with momentum ${K}=0$ (left panel)  and its radius in units $a=1$ (right panel) from  the 1D lattice model (red circles) with a one-orbital conduction band with $t_c=1$, and a $s,p_x$ two-orbital valence band with $t_{ss}=1.1, t_{pp}=-1, |t_{sp}|=0.5$. }The corresponding valence band structure is shown in Fig. \ref{fig:2}. Here, we assume $U_s=U_p=U$. The blue line shows the continuum approximation predictions.  The vertical dashed line marks the value of $U$ above which the continuum approximation starts to fail. See text for more details. 
}
\label{fig:dE_U2}
\end{figure}

\section{Results}
\label{sec:Results}
\subsection{The range of validity of the continuum approximation}

One way to validate the method described above is to verify that in the limit $U \rightarrow 0$, its results agree with those predicted by the continuum approximation. This will demonstrate that our method is also suitable for calculating the spectra of large excitons, although in this case the above-mentioned cutoff is large (it has to exceed the exciton radius) and the computation becomes expensive.

\begin{figure}[t]
\centering
\includegraphics[width=0.95\linewidth]{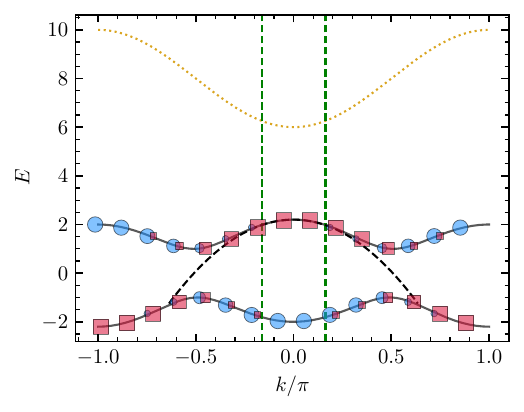} 
\caption{\color{black}{The conduction band (orange dotted line) and the two valence bands $E_{v}({k,\pm})$  (symbols)  for the parameters listed in Fig. \ref{fig:dE_U2}. The top/bottom of the valence/conduction band is at $k=0$, indicating a $K=0$ direct gap. The dashed black line shows the continuum approximation for the valence band, which matches the exact dispersion to within $1\%$ for $|k|\lesssim \pi/6$ (dashed vertical lines). The blue circles indicate the weight of $p_{x}$ orbital character while the red squares indicate the weight of $s$ orbital character.}}
\label{fig:2}
\end{figure}

Figure \ref{fig:dE_U2} compares the binding energy of the exciton with momentum ${K}=0$ and its corresponding exciton radius $\xi$ as predicted by the 1D lattice model with a $s,p_x$ two-orbital valence band (solid red circles) and by its corresponding continuum approximation (blue lines;  the relevant equations  are listed in Appendix \ref{App:1D_Cont}). For the lattice model, the  hopping integrals are set to $t_c= 1$, $t_{ss}=1.1, t_{pp}=-1$ and $|t_{sp}|=0.5$ (the signs due to the orbital overlaps are included in the Hamiltonian). For these values, the semiconductor has a direct gap, \textcolor{black} {which is why we focus on the lowest-energy exciton with ${K}=0$. The direct gap is demonstrated in Fig. \ref{fig:2}, which shows the dispersion of the conduction band $E_c(k) = - 2 t_c \cos(k)$ (dotted orange line) and of the two valence bands $E_{v}({k},\pm)$ in the Brillouin zone (symbols illustrate their orbital character, with blue/red associated with the $p_x$ and $s$ orbitals, respectively). The top of the top-valence band is at $k=0$. }

For these parameters, Fig. \ref{fig:dE_U2} shows that the continuum approximation is accurate for $U\le U_c \approx 2.2 t_c$, after which it becomes increasingly worse, see Fig. \ref{fig:dE_U2}. This corresponds to an exciton radius $\xi \gtrsim  2a$, which at first sight is a surprisingly low bound for a 'large' exciton. However, we can understand this as follows. For a given $\xi$, top-valence and conduction band states with momenta up to $k \sim 1/\xi$ contribute the most to the $K=0$ exciton $|K=0, {\rm exc}\rangle = \sum_{k} \phi_k c^\dagger_k v_{k,+}|{\rm GS}\rangle$. Here, $\xi \gtrsim 2a $ implies $|k|\lesssim {1\over 2a} \approx  {\pi \over 6a}$. This value is shown by the vertical dashed lines in Fig. \ref{fig:2}, and indeed provides a good bound for the region where the top-valence band states are described well by the continuum approximation (blue parabola). 

We have verified for multiple other parameter choices (not shown) that we always get a good estimate for the smallest exciton radius $\xi\sim 1/k_M$ for which a continuum approximation is still accurate, when we choose $k_M$ as the largest momentum for which a quadratic approximation is still reasonable near the top/bottom of the valence/conduction bands of that specific model. 

This simple procedure provides a practical recipe for estimating whether a continuum approximation is reasonable if the dispersions of the conduction and valence bands are known. As we point out in the next section, it may still be misleading in a multi-orbital system.

\subsection{Qualitative failure of the continuum approximation in multi-orbital systems}

In the example discussed above, we set $U_s=U_p$ for simplicity. However, in reality these quantities are highly dependent on the symmetry of the orbitals involved (their charge density) and their overlap with the conduction band orbital's charge density. In fact, there is no scenario that we can think of where $U_s=U_p$ is likely to hold.

Different Coulomb attractions for holes located in different orbitals add a second important ingredient in determining the exciton spectra, besides the dispersion of the top valence band in the multi-orbital system, discussed above. This is because a stronger Coulomb attraction for one of the orbitals may lead to a larger binding energy for excitons involving valence band states which have a maximal character of that type, even if these states are {\it not} at the top of the valence band. 

\begin{figure}[t]
\centering
\includegraphics[width=\linewidth]{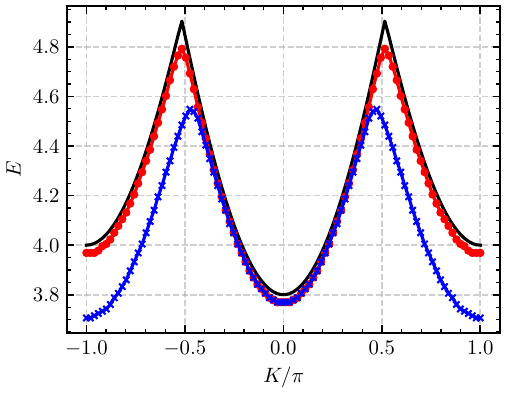}
\caption{Gap energy $E_{\rm gap}(K)$ (black solid line) for the 1D lattice model with parameters as in Fig. \ref{fig:dE_U2}. Red circles show the exciton dispersion $E_{\rm exc}(K)$ for  $U_p=U_s=0.5$, while blue crosses show $E_{\rm exc}(K)$ for  $U_p=3U_s=1.5$. The latter case stabilizes an exciton with momentum $K=\pi$ even though this is a direct gap ($K=0$) semiconductor, because the top-valence band states near $k=\pi$ have dominant $p$-character.}
\label{fig:UnevenConP_S}
\end{figure}

Such an example is illustrated in Fig. \ref{fig:UnevenConP_S}. The solid black line shows $E_{\rm gap}(K)$, {\it i.e.} the lower edge of the electron-hole continuum for different total momenta ${ K}$, for the same parameters as in Fig. \ref{fig:dE_U2}. Clearly, this is a direct-gap semiconductor with the minimum gap at $K=0$. Accordingly, the simplest continuum approximation predicts that the maximum exciton binding energy must appear at $K=0$. The exciton spectrum of the 1D lattice agrees with this prediction if $U_s=U_p= 0.5$, shown by the red circles. Indeed, here the strength of the attraction is the same independent of the character of the top-valence band, and therefore the lowest energy exciton appears where the electron-hole excitation gap is minimum. However, if we choose $U_p=3U_s= 1.5$ (blue crosses), then the exciton minimum {\color{black} transitions} to $K=\pm \pi$ because the top-valence band has a predominant $p$-character near $k=\pi$, see Fig. \ref{fig:2}. Note that the exciton dispersion near $K=0$ is little affected by the increase of $U_p$, consistent with the fact that near $k=0$ the top-valence band has primarily $s$-character in our model.

To summarize, if the continuum approximation simplifies the multi-orbital valence bands' dispersion to a single parabola centered at the top of the top-valence band, it can never reproduce the {\color{black} sharp transition} exemplified in this subsection, because information regarding the orbital character of the actual valence band is lost. Of course, \color{black} for Wannier-like excitons, one could proceed to a ``multi-valley" type of continuum approximation that would do better, although that removes the top advantage of the continuum approximation, namely its ability to produce analytical results. For small, Frenkel-like excitons, however, the bands need to be accurately described in the full Brillouin zone, and this simply cannot be achieved quantitatively with any quadratic-type of approximation, whether single- or multi-valley. A multi-valley continuum approximation might still be made to be qualitatively accurate even in this case, although we think that is less likely for the 2D example discussed next.  Nevertheless, we believe that our method makes it easier to compute the exact spectra of  small, Frenkel-like excitons than it would be to develop and validate a continuum approximation whose predictions might remain qualitatively correct in this limit.

\color{black}

	\subsection{2D exciton transition}

In this subsection, we illustrate another possible way for the continuum approximation to fail qualitatively by predicting the wrong momentum of the lowest energy exciton, in a multi-orbital model. 
Before proceeding, we mention that we verified that in the limit of weak electron-hole attraction, our method produces exciton results for this 2D model that are in agreement with those of the continuum approximation, see Appendices \ref{App:2D_Cont}-\ref{App:exp_break}.

\begin{figure}[t]
		\centering
	\includegraphics[width=\linewidth]{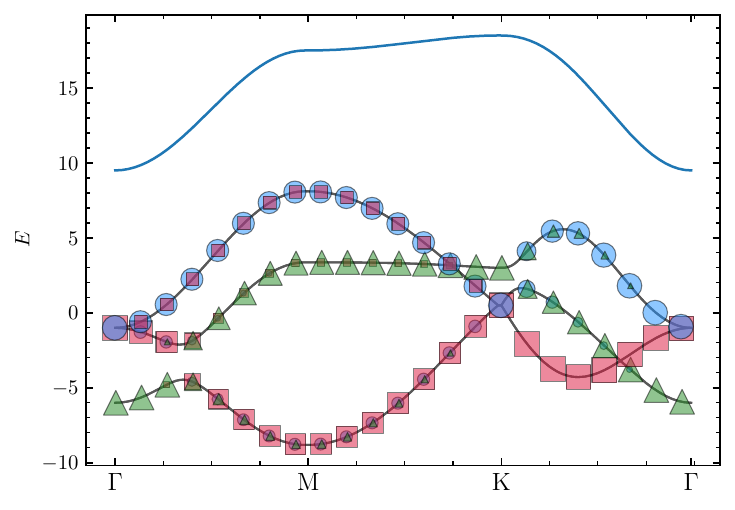}
		\caption{\color{black}Band structure for the 2D model. The conduction band is shown by the solid blue line. For the three valence bands, the blue circles indicate the weight of $d_{x^2-y^2}$ character, red squares indicate the $d_{xy}$ weight and green triangles indicate the $d_{3z^2-r^2}$ weight.  The valence band hopping parameters are listed in Appendix A and $t_c=1, \Delta = 15.5$. The band-gap is indirect, with the minimum along the $K-\Gamma$ direction, see next figure.}
		\label{fig:4}
	\end{figure}

As an example, we use the 2D triangular lattice with the three $d$ orbital bases introduced in Section II.  {\color{black}  Figure \ref{fig:4} shows the conduction band (blue solid line) and the three valence bands and their orbital character. Specifically, blue circles indicate the weight of the $d_{x^2-y^2}$ orbital, red squares indicate the $d_{xy}$ weight and green triangles indicate the $d_{3z^2-r^2}$ weight.  The choice of valence band hopping parameters is listed in Appendix A, and  we set  $t_c=1, \Delta = 15.5$. Symmetry guarantees that the overlaps of all $3d$ orbitals with the $s$-type conduction band orbital are the same, so we set $U_{\alpha}=U$ for all three valence orbitals. A short discussion of the effect of unequal $U_\alpha$ can be found in Appendix~\ref{App:OtherU}. }

\begin{figure}[t]
\centering
\includegraphics[width=\linewidth]{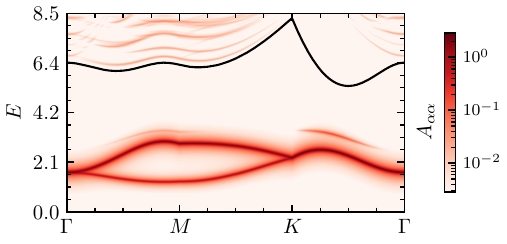}
\includegraphics[width=\linewidth]{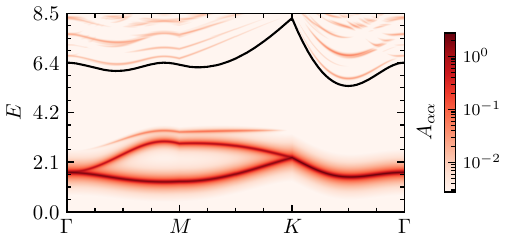}
\includegraphics[width=\linewidth]{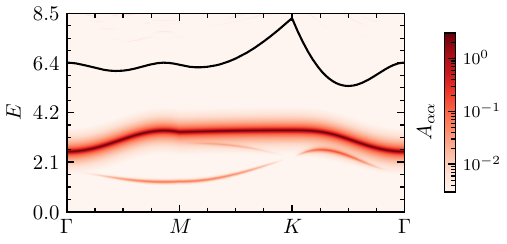}
\caption{Heat maps showing the spectral weight $A_{\alpha \alpha}({\bf K}, \omega) = - {1\over \pi} \text{Im}G_{\alpha \alpha} ({\bf K}, \bm{0}, z)$ for ${\bf K}$ along high-symmetry lines in the Brillouin zone, when $\alpha = d_{x^2-y^2}$ (top panel), $\alpha = d_{xy}$ (middle panel) and $\alpha = d_{3z^2-r^2}$ (bottom panel). Here $U=12$ and all other parameters are as listed in Appendix A. The superimposed thick black line marks the lower edge of the electron-hole continuum, $E_{\rm gap}({\bf K})$. See text for more details.}
\label{fig:5}
\end{figure}

Figure \ref{fig:5} shows the spectral weight $A_{\alpha \alpha}({\bf K}, \omega) = - {1\over \pi} \text{Im} G_{\alpha \alpha} ({\bf K}, \bm{0}, z)$ for ${\bf K}$ along high-symmetry lines in the Brillouin zone, for $\alpha = d_{x^2-y^2}$ (top panel), $\alpha = d_{xy}$ (middle panel) and $\alpha = d_{3z^2-r^2}$ (bottom panel). The thick black line marks the lower edge of the electron-hole continuum, $E_{\rm gap}({\bf K})$, \color{black} with an indirect minimum gap lying nearly half-way between $K$ and $\Gamma$. Above this edge, the spectral function contains continuum states, but their weight is too small to be clearly visible on the color scale used.
The relevant features are the three exciton bands visible well inside the gap (because of the large $U$ value used). These are discrete states, but here are broadened by the finite $\eta$ value we used. \color{black}In the $d_{3z^2-r^2}$ projection (bottom panel) most of the intensity resides in the highest-energy ({\em i.e.} lowest binding energy) exciton branch. Along the $\Gamma-M-K$ cuts, the lowest energy exciton branch appears with comparable intensity in the top two panels, showing that it has comparable amounts of $d_{x^2-y^2}$ and $d_{xy}$ character for these momenta. By contrast, along the $K-\Gamma$ direction, the top panel shows two exciton branches while the middle panel shows a third, different branch. Clearly, along this cut there is no mixing between these two orbitals, and the excitons have either $d_{x^2-y^2}$ or $d_{xy}$ character.

The simplest continuum approximation would predict that the lowest energy exciton appears along the $K-\Gamma$ line, at the momentum ${\bf K}$ for which $E_{\rm gap}({\bf K})$ has the global minimum.  The middle panel indicates that the lowest energy exciton branch has indeed a local minimum at that momentum. However, a more careful inspection shows that the global minimum is actually at the $M$ point. We note that the $M$ point is not even a local minimum for $E_{\rm gap}({\bf K})$!

%\newpage
%	There is a qualitative shift for the ground state exciton as \(U\) increases, as depicted by Figs. \ref{fig:2D_3band_exciton_contour_scan} and \ref{fig:2D_3band_exceiton_transition}. (CHOOSE which of these work better, separate or same plot.) (The detailed change of the energy contour can be found in Appendix \ref{App:3Band_kapace}).

	\begin{figure}[t]
		\centering
	\includegraphics[width=\linewidth]{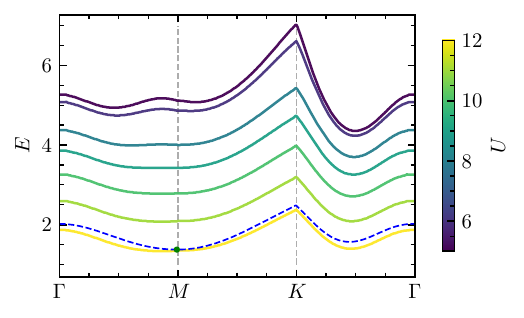}
	\includegraphics[width=\linewidth]{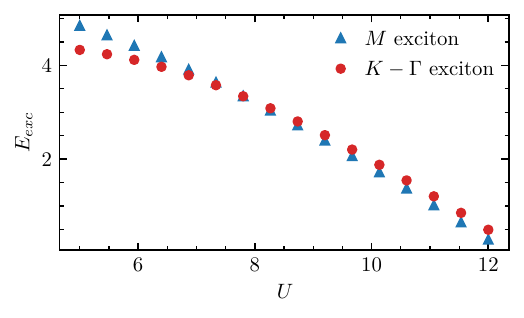}
		\caption{(top) Dispersion of the lowest-energy exciton $E_{\rm exc}({\bf K})$ for different values of $U$. The blue dashed line is the exciton energy calculated with a cut-off at $\delta_M =2$ (see Appendix \ref{App:1shell} for details), and the dot marks its predicted lowest-energy exciton; (bottom) $E_{\rm exc}({\bf K})$ when ${\bf K}=M$ (blue triangles) and when ${\bf K}$ is the indirect gap, {\it i.e.} local minimum for $E_{\rm gap}({\bf K})$ along the $K-\Gamma$ line (red circles).  }
		\label{fig:6}
	\end{figure}

To understand what happens, we plot in the top panel of Fig. \ref{fig:6} the energy of the lowest-energy exciton branch along the same cuts, for increasing values of $U$. As expected, the exciton moves to lower energies as the attraction $U$ increases. Its dispersion closely follows the gap energy $E_{\rm gap}({\bf K})$ for small values of $U$, however, this changes as $U$ increases. 
%{\bf At 
%$U=12$, a nearest-neighbor (‘one-shell’) calculation already reproduces the qualitative features (see Appendix \ref{App:1shell} for detailed calculation). The green dot marks the ground-state momentum predicted by that approximation.}

\begin{figure}[t]
\includegraphics[width=0.8\linewidth]{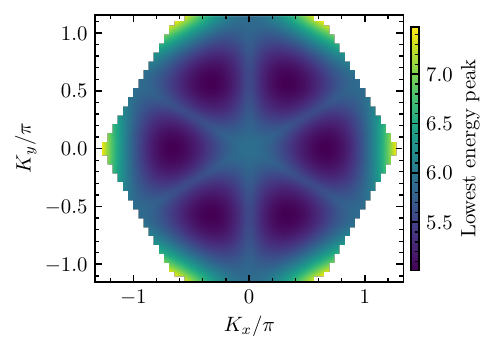}
\includegraphics[width=0.8\linewidth]{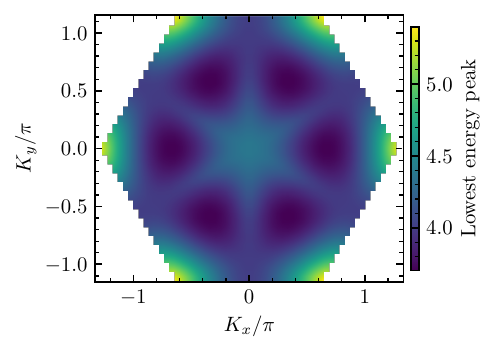}
\includegraphics[width=0.8\linewidth]{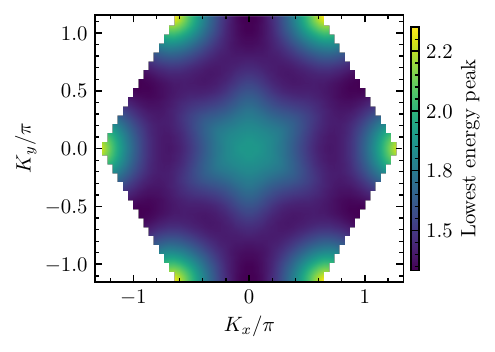}
	\caption{Contour plots of the lowest-energy exciton band  $E_{\rm exc}({\bf K})$ inside the Brillouin zone, for $U=2.5$ (top), $U=8$ (middle) and $U=12$ (bottom).}
	\label{fig:7}
\end{figure}

\begin{figure}[t]
\includegraphics[width=0.8\linewidth]{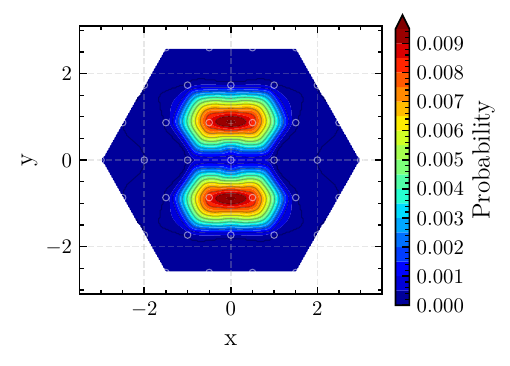}
\includegraphics[width=0.8\linewidth]{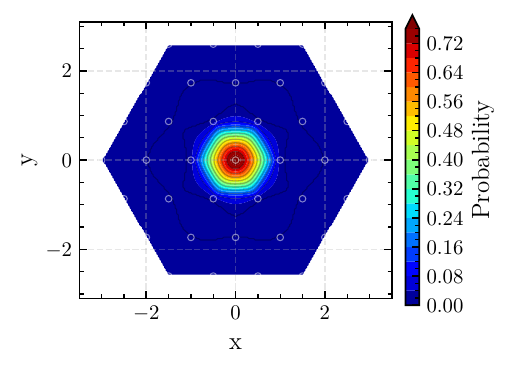}
\includegraphics[width=0.8\linewidth]{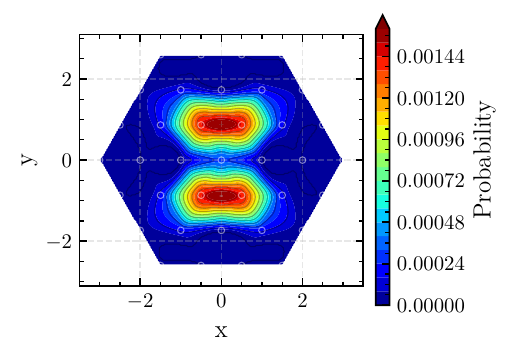}
\caption{\textcolor{black}{Probability $|\phi_{\bf K}(\beta, \bm{\delta})|^2$
plotted as a function of the displacement $\bm{\delta}$ between electron and hole, for the lowest-energy exciton and for ${\bf K}$ equal to the indirect gap momentum, and $U=12$. The projections are on orbital  $\beta = d_{x^2-y^2}$ (top), $\beta = d_{xy}$ (middle) and $\beta=d_{3z^2-r^2}$ (bottom). The triangular lattice sites are shown by the white circles. The probability is defined only at these lattice points; the interpolated continuous heat map is shown to guide the eye. }}
	\label{fig:8}
\end{figure}

The bottom panel tracks the exciton energy at the momentum $M$ and at the indirect gap momentum lying on the $K-\Gamma$ line. The two curves cross at $U\approx 8$, marking a {\color{black}sharp transition} in the momentum and character of the lowest-energy exciton. This is another example of a sharp transition that would be missed by a continuum approximation.  

This change in the shape of the dispersion of the lowest-energy exciton band is illustrated graphically in Fig. \ref{fig:7}, where its energy $E_{\rm exc}({\bf K})$ is plotted in the full Brillouin zone. For the smaller $U=2.5$ value (top panel) its minimum is clearly along the $\Gamma-K$ cut. However, as $U$ increases, the exciton energy at the $M$ point drops faster so that for $U=12$ (bottom panel) the minimum is at the $M$ point. 

\begin{figure}[t]
\includegraphics[width=0.8\linewidth]{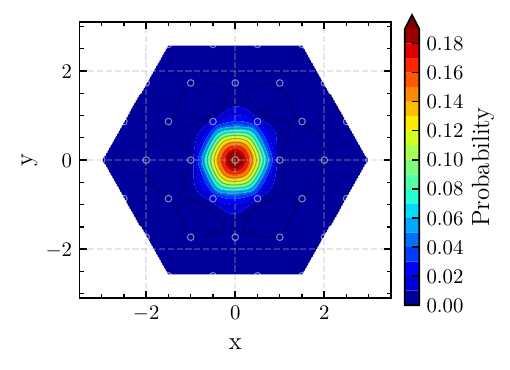}
\includegraphics[width=0.8\linewidth]{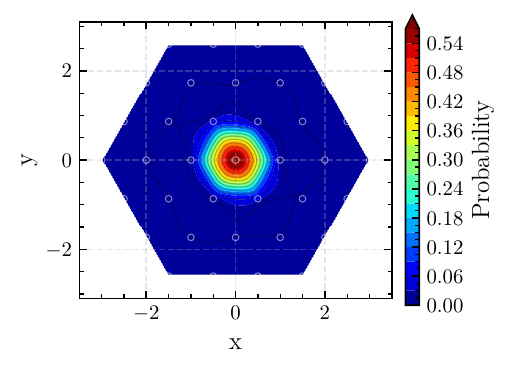}
\includegraphics[width=0.8\linewidth]{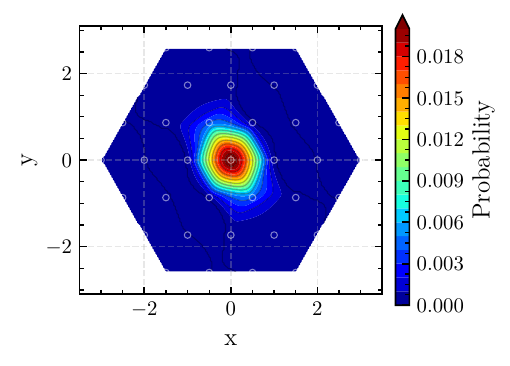}
\caption{Same as in Fig. \ref{fig:8} but for ${\bf K}=M$.
	}
	\label{fig:9}
\end{figure}
This transition can be attributed to the symmetry already mentioned, which prevents mixing of $d_{x^2-y^2}$ and $d_{xy}$ orbitals along the $\Gamma-K$ line. \color{black} This is supported by the plots of the real-space probability $|\phi_{\bf K}(\beta, \bm{\delta})|^2$ to find the hole in orbital $\beta$ and at a displacement $\bm{\delta}$ from the electron, shown in Fig. \ref{fig:8} when ${\bf K}$ is the indirect band-gap momentum, and in Fig. \ref{fig:9} when ${\bf K}= M$ point. \color{black}

First, both of these excitons are Frenkel-like, with most of the probability concentrated at $\bm{\delta}=0$. The exceptions are the $d_{x^2-y^2}$ and the $d_{3z^2-r^2}$ orbitals (top and bottom panel in Fig. \ref{fig:8}, respectively) where symmetry prevents weight at $\bm{\delta}=0$ and instead there is a very small amount of weight  at $|{\bm \delta}|=a$. This is consistent with the vanishing spectral weight of this exciton in the top plot of Fig. \ref{fig:5}. By contrast, the $M$ point exciton has considerable on-site probability in all orbitals, see Fig. (\ref{fig:9}). 
The reason why the M exciton lowers its energy faster with increasing $U$ is now clear: this exciton has the larger total probability of displacement $\bm{\delta}=0$ between the electron and hole, and therefore experiences more of the on-site $U$ attraction. 

Such a discontinuous transition in the momentum of the lowest-energy exciton would again be completely missed by a continuum approximation. 
We note that the argument above can be made more quantitative by using a very simple variational approximation, where we allow this exciton to spread only over displacements $|\bm{\delta}|\le 1$, {\em i.e.} we set $\delta_M=2$. The corresponding predicted exciton energy when $U=12$ is shown as the dashed blue line in Fig. \ref{fig:6}(top). It is qualitatively correct, with the minimum energy (shown by the dot) in excellent agreement with the converged calculation, which has $\delta_M=10$. The quantitative differences are because at these ${\bm K}$ values, there is still some weight on states with $|{\bm \delta}|\ge 2$. Still, this shows that a very simple calculation involving only 7 sites (in this case) already captures the essential behavior of this {\color{black} small, Frenkel-like excitons}. Details of this minimal calculation and further insights that can be obtained from it are presented in Appendix \ref{App:1shell}.

	%\begin{figure}[htbp]
	%	\centering
	%	\includegraphics[width=\linewidth]{2D_3band/2D_3band_exciton_contour_scan.pdf}
	%	\caption{Exciton ground state dispersion morphology changes as \(U/t\) changes. The darker, blue curve is when $U/t$ is small and it agrees with the continuum model that the exciton ground state will be on the \(K-\Gamma\) line. However, as the $U/t$ increases, indicated by the lighter yellow line, the ground state changed to near \(M\) point.}
	%	\label{fig:2D_3band_exciton_contour_scan}
	%\end{figure}
	
	%\begin{figure}[htbp]
	%	\centering
	%	\includegraphics[width=\linewidth]{2D_3band/2D_3band_exciton_transition.pdf}
	%	\caption{A closer look to the exciton energy as $U/t$ changes. The exciton energy local minimum on the $K-\Gamma$ line is plotted as red dots with the exciton energy local minimum of the $M$ which is in blue triangles. Exciton ground state energy transits from minimum along the \(K-\Gamma\) line to \(M\).}
	%	\label{fig:2D_3band_exciton_transition}
	%\end{figure}
	
%	\paragraph*{Weak coupling.}
%	For $|U|\ll \text{bandwidth} \approx 10t$ a single bound branch emerges just below the electron-hole continuum with its minimum on the $K$–$\Gamma$ line.
	
%	\paragraph*{Strong coupling.}
Finally, while the discussion was focused on the lowest-energy exciton and the transition with $U$ in its character and momentum, our results also show that as $U$ grows, additional exciton branches detach from the continuum, such that there are 3 branches for $U=12$. Their energies, orbital character and real-space wavefunctions (or probabilities) can be calculated similarly.

	\section{Conclusions}
	\label{sec:conclusion}

\color{black}
In this work, we adopted a real-space approach  to study exciton spectra and their wavefunctions in multi-orbital lattice models. The method can be used for excitons of any size, from Wannier-like to Frenkel-like. However, it becomes computationally most efficient as one approaches the limit of small, Frenkel-like excitons. 

For the large Wannier excitons that appear in the limit of weak electron-hole attraction, we verified that our lattice results agree with the prediction of the corresponding continuum approximation. This comparison allowed us to propose a simple criterion to estimate when the continuum approximation is likely to fail quantitatively in models with a single valence and a single conduction band.

It is well known that continuum approximations must  fail quantitatively, eventually, when applied to models that have small, Frenkel-like excitons because the full details of the bandstructure are relevant for these excitons' spectra. Here, we showed that the simplest continuum approximations can also fail qualitatively in  multi-orbital models. Specifically, we presented two models where, in certain regions of the parameter space, the momentum of the lowest-energy exciton is not set by the momentum of the gap (whether the gap is direct or indirect).  The simplest continuum approximations expand the valence/conduction bands about their maximum/minimum values, so by construction they anticipate that the lowest-energy exciton's momentum  equals the momentum of the gap. More involved continuum approximations can be envisioned and these might be able to predict the qualitatively correct momentum, but we believe that for small, Frenkel-like excitons, it is easier to find the exact lattice spectrum with the method proposed here.

These examples are just beginning to explore possible new behavior of small, Frenkel-like excitons in models with multi-orbital bands. For such models, our real-space method provides a very efficient way to calculate the exciton spectra and their wavefunctions in the full Brillouin zone. Furthermore, the method can be extended to a few more carriers \cite{MB11}, relevant for calculating trions and spectra of larger complexes, to inclusion of couplings to bosonic degrees of freedom such as phonons or magnons, and to addition of interfaces and/or defects in the model. We believe that this makes it a valuable tool to explore various model Hamiltonians to understand possible behavior of small, Frenkel-like excitons.

\color{black}
	
	\section{Acknowledgments}
	
This project was undertaken thanks in part
to funding from the Max Planck-UBC-UTokyo Center for Quantum Materials and the Canada First Research Excellence Fund, Quantum Materials and Future Technologies Program, as well as the Natural Sciences and Engineering Research
Council of Canada (NSERC). We gratefully acknowledge the use of computing resources from the Stewart Blusson Quantum Matter Institute computing cluster LISA. Lastly, we also acknowledge that the research is done on the UBC Point Grey (Vancouver) campus, located on the traditional, ancestral, unceded territory of the $x^wm$\textipa{@}$\theta k^w$\textipa{@}$\overset{,}{y}$\textipa{@}$m$ (Musqueam) First Nation.

		\appendix
	\section{Hoppings in the 2D, three-orbital valence band}
	\label{App:hopping}
    We work in the orbital basis $(d_{x^2-y^2},\,d_{xy},\,d_{3z^2-r^2})$. We choose these \(d\) orbitals as an example and for their non-trivial symmetry with respect to the underlying lattice; any other basis can be treated similarly. Following Harrison's rules and the Slater-Koster matrix elements \cite{Harrison1989,Slater1954}, the nearest-neighbor hopping integrals for a bond parallel to $\hat x$ are parameterized as follows: if the $d_{3z^2-r^2}\leftrightarrow d_{3z^2-r^2}$ hopping is denoted $ t_{v3}$, then the $d_{x^2-y^2}\leftrightarrow d_{x^2-y^2}$ hopping is $t_{v1}/t_{v3}=19/9$,  the $d_{xy}\leftrightarrow d_{xy}$ hopping is $t_{v2}/t_{v3}=-16/9$, and the $d_{x^2-y^2}\leftrightarrow d_{3z^2-r^2}$ hopping $t_{v4}/t_{v3}=-5\sqrt{3}/9$; all other hopping integrals are zero for symmetry reasons.

    For example, 
    $
    t_{v1}=\tfrac{3}{4}V_{dd\sigma}+\tfrac{1}{4}V_{dd\delta}$, $
    t_{v3}=\tfrac{1}{4}V_{dd\sigma}+\tfrac{3}{4}V_{dd\delta}$ so $
    {t_{v1}}/{t_{v3}}=(3r+1)/(r+3)$ where $r={V_{dd\sigma}}/{V_{dd\delta}}$ and  $V_{dd\sigma}$ and $V_{dd\delta}$ are the two-center Slater--Koster $dd\sigma$ and $dd\delta$ matrix elements between neighboring $d$ orbitals.
    Using Harrison’s estimate $r= V_{dd\sigma}/V_{dd\delta}\approx (-6)/(-1)$ gives $t_{v1}/t_{v3}=19/9$. The other ratios are obtained similarly.

    In matrix form, the hoppings along the bond ${\bm \delta}_0= \hat x$ are, thus:
	\begin{equation}
		t_{\alpha \beta} ({\boldsymbol \delta_{0}}) \;=\; \begin{pmatrix} t_{v1} & 0 & t_{v4} \\ 0 & t_{v2} & 0 \\ t_{v4} & 0 & t_{v3} \end{pmatrix}.
		\label{eq:T_ref}
	\end{equation}
	The hopping matrix for a bond  $ {\boldsymbol \delta_{b}}$ oriented at an angle $\theta$ from $\hat x$  are generated by  rotations in the $(x,y)$ plane:
    \begin{align}
        t_{\alpha \beta} ({\boldsymbol \delta_{b}})\;=\;\underline R^{\!\top}(\theta)\,t_{\alpha \beta} ({\boldsymbol \delta_{0}})\,\underline R(\theta).
    \end{align}
    where
	\begin{align}
		\underline R(\theta)\;=\; \begin{pmatrix} \cos 2\theta & \;\sin 2\theta & 0 \\ -\sin 2\theta& \;\cos 2\theta & 0 \\ 0&0&1 \end{pmatrix}
        \label{eq:tab}
	\end{align}
    In a triangular lattice, there are three inequivalent directions
    $\theta\in\{0^\circ,60^\circ,120^\circ\}$ corresponding to the three bond
    vectors
    $\boldsymbol\delta_0=(1,0)$,
    $\boldsymbol\delta_1=(\tfrac12,\tfrac{\sqrt3}{2})$,
    $\boldsymbol\delta_2=(-\tfrac12,\tfrac{\sqrt3}{2})$;
    the remaining three nearest-neighbors are their opposites
    $-\boldsymbol\delta_b$ ($b=0,1,2$).

    For the specific calculations shown in the main paper, we set $t_{v3}=1$, the electron hopping in the conduction band is $t_c=1$, and $\Delta = 15.5$.

    \section{Recurrence relations and their convergence}
    \label{sec:Conv_1D}

    \subsection{1D chain formalism}
    The  1D chain Hamiltonian with a multi-orbital valence band is given in Eqs.~(\ref{2})--(\ref{6}). We work in the exciton basis 
    \(|\alpha, K, \delta\rangle\), where \(K\) is the center-of-mass crystal momentum, \(\delta\in\mathbb{Z}\) is the electron--hole separation (in lattice spacings), and \(\alpha\in\{s,p\}\) indexes the valence orbital. For a fixed source orbital \(\gamma\in\{s,p\}\), we collect the two orbital components of the resolvent into the vector Green's function
    \begin{equation}
        \bm G_{\gamma,\delta}(K,z) \equiv 
    \begin{bmatrix}
    G_{\gamma,s}(K,\delta,z)\\[2pt]
    G_{\gamma,p}(K,\delta,z)
    \end{bmatrix}.
    \end{equation}

    We start by operating the Hamiltonian \(H\) on a basis state defined in Eq. \ref{7}. 
    For \(\delta\neq 0\), the on-site part of \(H\)  leaves $\delta$ unchanged, while hopping changes \(\delta\to\delta\pm 1\).  
    %Then we adjust the phase of the basis \(\delta\to\delta\pm 1\) so that it carries the correct center-of-mass phases of the basis \(\delta\pm 1\). This creates an additional phase with momentum \(K\) on the basis, yielding \(e^{\pm iK/2}\). Hoppings of electrons in the conduction band have negative phases while the valence band hoppings give positive phases. \\
 %   \textbf{Example}
%   \begin{align}
 %   &c_{n+1}^\dagger c_n|\alpha, K, \delta\rangle \nonumber\\
 %   &= \frac{1}{\sqrt{N}}\sum_{n} e^{iK(R_n+\delta/2)} c_{n+1}^\dagger c_n c_n^\dagger v_{\alpha,n+\delta}|\text{GS}\rangle \nonumber\\
 %   &= \frac{1}{\sqrt{N}}\sum_{n} e^{iK(R_n+\delta/2)} c_{n+1}^\dagger v_{\alpha,n+\delta}|\text{GS}\rangle\nonumber\\
 %   &= \frac{1}{\sqrt{N}}\sum_{m} e^{iK(R_m-1+\delta/2)} c_{m}^\dagger v_{\alpha,m+\delta-1}|\text{GS}\rangle \nonumber\\
%    &= e^{-iK/2}\frac{1}{\sqrt{N}}\sum_{m} e^{iK(R_m+(\delta-1)/2)} \nonumber\\
%    &\qquad\qquad\times c_{m}^\dagger v_{\alpha,m+\delta-1}|\text{GS}\rangle \nonumber\\
%    &= e^{-iK/2}|\alpha, K, \delta-1\rangle.
%    \end{align}    

For the  \(s\)-component, Eq. (\ref{13}) then becomes:
    \begin{align}
        &(z-\Delta)\,G_{\gamma,s}(K,\delta,z) = \nonumber\\
        &\quad \big(t_{ss} e^{-iK/2}-t_c e^{+iK/2}\big)\,G_{\gamma,s}(K,\delta{+}1,z) \nonumber\\
        &\quad+ t_{sp} e^{-iK/2}\,G_{\gamma,p}(K,\delta{+}1,z) \nonumber\\
        &\quad+ \big(t_{ss} e^{+iK/2}-t_c e^{-iK/2}\big)\,G_{\gamma,s}(K,\delta{-}1,z) \nonumber\\
        &\quad+ t_{sp} e^{+iK/2}\,G_{\gamma,p}(K,\delta{-}1,z),
        \label{eq:example_action}
    \end{align}
    The equation for $G_{\gamma,p}(K,\delta,z)$ looks similar but  \(t_{ss}\to t_{pp}\).
    
    We then group them in the recurrence relation:
    \begin{equation}
    \underline\alpha\,\bm G_{\gamma,\delta}(K,z) \;=\; \underline\beta\,\bm G_{\delta+1}(K,z) \;+\; \underline\gamma\,\bm G_{\delta-1}(K,z),
    \label{eq:bulk_recurrence}
    \end{equation}
    where for $\delta \ne 0$ the $2\times 2$ matrices are:
    \begin{align}
    \underline \alpha &= 
    \begin{bmatrix} z - \Delta & 0 \\[2pt] 0 & z - \Delta \end{bmatrix}, \\
    \underline \beta &= 
    \begin{bmatrix} t_{ss} e^{-iK/2} - t_c e^{+iK/2} & t_{sp} e^{-iK/2} \\[2pt]
    t_{sp} e^{-iK/2} & t_{pp} e^{-iK/2} - t_c e^{+iK/2} \end{bmatrix}, \\
    \underline \gamma &= 
    \begin{bmatrix} t_{ss} e^{+iK/2} - t_c e^{-iK/2} & t_{sp} e^{+iK/2} \\[2pt]
    t_{sp} e^{+iK/2} & t_{pp} e^{+iK/2} - t_c e^{-iK/2} \end{bmatrix}.
    \end{align}

These are solved in terms of the ansatz $\bm G_{\gamma,\delta}(K,z)= \underline{S}_{\pm}\bm G_{\gamma,\delta\mp 1}(K,z)  $, where the upper/lower signs correspond to $\delta>0$ and $\delta <0$, respectively, and $ \underline{S}_+= [\underline{\alpha} -\underline{\beta} \underline{S}_+]^{-1} \underline{\gamma} $ and $ \underline{S}_-= [\underline{\alpha} -\underline{\gamma} \underline{S}_-]^{-1} \underline{\beta} $ are continued fractions of matrices, calculated with $M$ iterations starting from ${\underline S}_\pm =0$. The cutoff $M$ is chosen sufficiently large to achieve convergence to the desired accuracy (see below). Physically, $M$ defines a variational solution whereby the electron and hole are not allowed to be $M$ or more sites apart, {\em i.e.} setting to zero  all propagators with $|\delta|\ge M$. \textcolor{black}{We note that in the main text, we called this cut-off $\delta_M$} instead of $M$.  
    
    The explicit form of Eq. (\ref{13}) for $\delta=0$ reads:
    \begin{equation}
(\underline\alpha+\underline{\mathcal {U}})\,\bm G_{\gamma,0} \;=\; \mathbf e_{\gamma} \;+\; \underline\beta\,\bm G_{\gamma,1} \;+\; \underline\gamma\,\bm G_{\gamma,-1},
    \label{eq:boundary}
    \end{equation}
    where \(\underline{\mathcal {U}} = \mathrm{diag}(U_{s},U_{p})\), \(\mathbf e_{\gamma}=\begin{bmatrix}\delta_{\gamma s}\\[2pt]\delta_{\gamma p}\end{bmatrix}\). It is solved by using $\bm G_{\gamma,\pm}= \underline{S}_{\pm} \bm G_{\gamma,0}$ to find:
    \begin{equation}
    \bm G_{\gamma,0}(K,0,z) \;=\; \big[\,\underline\alpha+\underline{\mathcal {U}} - \underline\beta\, \underline S_+ - \underline\gamma\, \underline S_-\,\big]^{-1}\,\mathbf e_{\gamma}.
    \label{eq:green_solution}
    \end{equation}
    From this we can then get any $\bm G_{\gamma,\delta} =  {\underline S}_\pm^{|\delta|}\bm G_{\gamma,0}$, depending on whether $\delta$ is positive or negative.

\subsection{2D lattice formalism}

    The method described above can be generalized straightforwardly to lattices in higher dimensions \cite{BerciuCookEPL2010}. Here we briefly describe the approach for a triangular lattice. The relative displacement between electron and hole is $\boldsymbol\delta\in\mathbb{Z}^2$ and we define $N=|\boldsymbol\delta| $ (in units of $a$) as the minimum number of hops to get from an onsite configuration  ($\delta =0$) to the relative separation $\boldsymbol\delta$.  For a triangular lattice there are $6N$ distinct ${\bm \delta}$ separations for a given $N$, lying on a hexagonal shell. We index these  counter-clockwise starting from the one lying along the $+x$-axis, and group together into $\bm G_{\gamma, N}$ the $d_N =18N$ Green's functions with the same $N$. \textcolor{black}{The extra factor of 3 is because in our model we consider  3 valence orbitals and 1 conduction orbital per site; generalization to other numbers of valence and/or conduction orbitals is straightforward.}

    In models with nearest-neighbor hopping, the equation of motion Eq. (\ref{13}) for any Green's function with a given $N$ only links it to other Green's functions with the same $N$ or with $N\pm 1$. As a result, they can be again combined into a matrix recurrence relation, where for any $N\ge 1$:
    \begin{align}
    \underline\alpha_N(\mathbf K,z)\,\bm G_{\gamma, N}
    =   \underline\beta_N(\mathbf K)\,\bm G_{\gamma, N+1} 
     + \underline\gamma_N(\mathbf K)\,\bm G_{\gamma, N-1},
    \label{B8}
    \end{align}
 The matrices 
    $\underline\alpha_N\in\mathbb{C}^{d_N\times d_N}$,
    $\underline\beta_N\in\mathbb{C}^{d_N\times d_{N+1}}$,
    $\underline\gamma_N\in\mathbb{C}^{d_N\times d_{N-1}}$ can be read off from Eq. (\ref{13}). Eq. (\ref{B8}) is solved using the ansatz $\bm G_{\gamma, N}= {\underline S}_N (\mathbf K,z) \bm G_{\gamma, N-1}$, where  ${\underline S}_N(\mathbf K,z)= [  \underline\alpha_N(\mathbf K,z) - \underline\beta_N(\mathbf K) {\underline S}_{N+1}(\mathbf K,z)]^{-1}\underline\gamma_N(\mathbf K)$ is calculated iteratively starting from $\underline S_{M}(\mathbf K,z)=0$ for a sufficiently large cutoff $M$ such that the desired accuracy is reached (see example below). In particular, this allows us to calculate $\underline S_1(\mathbf K,z)$.  \textcolor{black}{For clarity, we note that $M$ is related to the generic cutoff $\delta_M$ used in the main text; the explicit relation depends on the lattice symmetry}.
    
    The equation for $N=0$ reads:
    \begin{align}
    (\underline\alpha_0+\underline{\mathcal {U}})\bm G_{\gamma, 0} &= \mathbf e_{\gamma}+\underline\beta_0(\mathbf K)\,\bm G_{\gamma, 1}
    \label{B9}
    \end{align}
    where
    \begin{align}
    \underline{\mathcal {U}} &= \mathrm{diag}(U_{x^2-y^2},U_{xy},U_{3z^2-r^2})
    \label{eq:2D_boundary_alpha0}
    \end{align}
    and the $\mathbf e_{\gamma}$ vector has an entry 1 for orbital $\gamma$, and 0 for the other two orbitals.
    \begin{figure}[t]
    	\centering
    	\includegraphics[width=\linewidth]{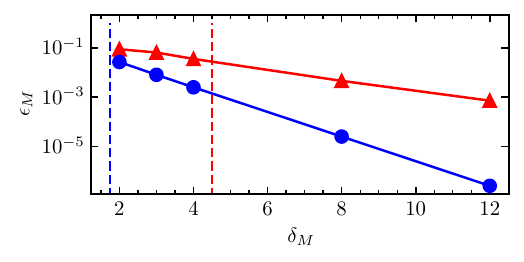}
        \includegraphics[width=\linewidth]{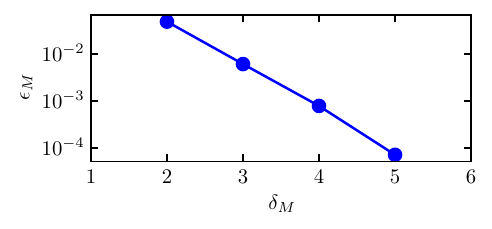}
    	\caption{Relative error $\epsilon_M=|E_{\rm exc}(M)/ E_{\rm final}-1|$ as a function of $M$, where $E_{\text{exc}}(M)$  is the  lowest exciton energy calculated using truncation at  $M$,  while \(E_{\text{final}}\) is the converged value for the lowest exciton energy,  obtained by truncating  at a sufficiently large cutoff $M$ such that increasing it by 1 results in a relative change below $10^{-6}$.
        (a) 1D results: blue circles correspond to a large  $U_p = U_s = 2.2$, while red triangles correspond to a weaker $U_p = U_s = 0.8$. In both cases $t_c = 1, t_{ss} = 1.1, t_{sp} = 0.5, t_{pp} = -1.0$. The former case has a smaller exciton (the dashed vertical lines show the corresponding exciton radius $\xi$) and thus converges faster with increasing $M$; (b) corresponding 2D results for the three band model with $U_{x^2-y^2}=U_{xy}=U_{3z^2-r^2}= 12$ and the other parameters listed in Appendix~\ref{App:hopping}. Here \(\xi \approx 0.856\) (not shown) is very small and the convergence with $M$ is very rapid.}
    	\label{fig:1D_bind_E_conv}
    \end{figure}
    
    The solution of Eq. (\ref{B9}) is, then:
    \begin{equation}
    \bm G_{\gamma, 0}(\mathbf K,z) = \Big(\underline\alpha_0+\underline{\mathcal {U}} -\underline\beta_0\,{\underline S}_1(\mathbf K,z)\Big)^{-1}\mathbf e_{\gamma}.
    \end{equation}
    Truncating at shell radius $N = 1$  ({\em i.e.} setting $M=2$ when calculating the continued fractions $\underline S_N$), gives the one-shell approximation used in Appendix~\ref{App:1shell}. The cut-off used for  Fig.~\ref{fig:5}-\ref{fig:9} in the main text is \(M=10\).

The convergence in terms of the cutoff $M$ is illustrated in Fig. \ref{fig:1D_bind_E_conv} for sample  1D (top) and 2D (bottom) cases. The stronger the attraction and therefore the smaller the exciton, the faster the convergence with increasing $M$. For really small, Frenkel-like excitons, even a cutoff $M=2$ can already give reasonable accuracy, as further discussed in Appendix~\ref{App:1shell}. This illustrates the main advantage of using this method for strongly bound excitons.

    \section{Longer-range Coulomb attraction}
    \label{App:OtherU}

Inclusion of longer-range Coulomb interactions is trivial in this method: such terms are added to the corresponding $\underline \alpha$ matrix, similar to what was done above for $N=0$. As a result, the computational cost is the same as for an on-site attraction, if the resulting excitons have comparable radii.

    \begin{figure}[t]
    	\centering
    	\includegraphics[width=\linewidth]{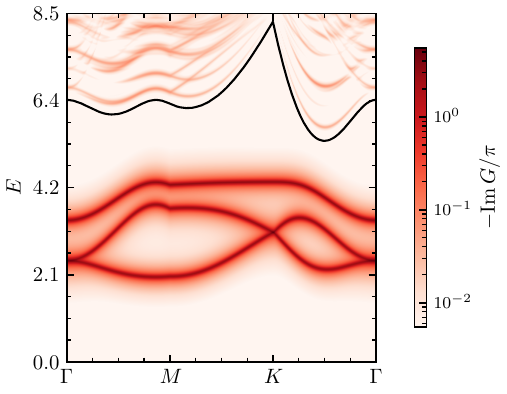}
    	\caption{Heat map similar to Fig.~\ref{fig:4} showing the spectral weight $A_{\alpha \alpha}({\bf K}, \omega) = - \frac{1}{\pi} \text{Im}G_{\alpha \alpha} ({\bf K}, \bm{0}, z)$ for ${\bf K}$ along high-symmetry lines in the Brillouin zone, summing over all the source orbitals. Parameters are: on-site attraction $U=11$, nearest-neighbor attraction $U_1=2$, all other parameters are as defined in Appendix~\ref{App:hopping}.}
    	\label{fig:U_1}
    \end{figure}

    As an example, in Fig. \ref{fig:U_1} we show the exciton spectrum when a nearest neighbor attraction $U_1=2$ is included beside the on-site $U=11$; all other parameters are as listed in  Appendix~\ref{App:hopping}. 
 Importantly, we find that adding a nearest-neighbor attraction preserves  the sharp ground-state momentum transition discussed in the main text, and in fact it provides additional tunability to achieve it. Arbitrarily long-range attraction can be added and studied similarly. However, very small excitons only "sample" short relative displacements, so whether there is attraction at displacements larger than the cutoff $M$ is irrelevant.

\begin{figure}[t]
    	\centering
    	\includegraphics[width=\linewidth]{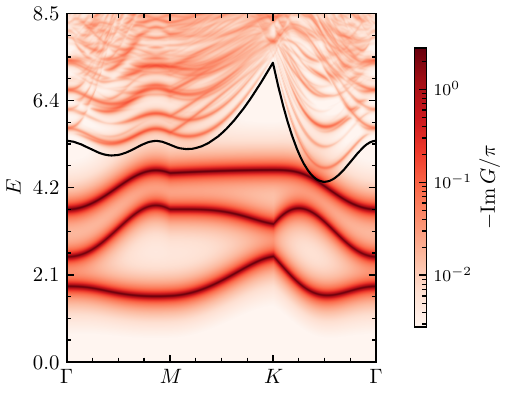}
    	\caption{Similar heat map as in Fig.~\ref{fig:U_1}, but now for on-site attraction \(U_{x^2-y^2} = 9.5,\,U_{xy} = 10.5,\,U_{3z^2-r^2} = 9.5\) and nearest-neighbor \({ U}_1=0\). }
    	\label{fig:Unequal_U}
    \end{figure}

Orbital-dependent on-site values $U_\alpha$ for the Coulomb attraction  between the electron and the hole in orbital $\alpha$ can be considered similarly, by setting $\underline{\mathcal U}=\mathrm{diag}(U_{\alpha})$.
In the main text we considered the case $U_{\alpha}=U$ to highlight the sharp transition in the ground state.
    Allowing orbital anisotropy $U_{\alpha}$ will change the results quantitatively, but the sharp transition persists except for highly fine-tuned parameters where the $U_\alpha$ vary very significantly with $\alpha$.  An example plot of the exciton spectrum  for \(U_{x^2-y^2} = 9.5,\,U_{xy} = 10.5,\,U_{3z^2-r^2} = 9.5\) is shown in Fig. \ref{fig:Unequal_U}.

Generalizations to orbital-dependent longer-range Coulomb attraction and/or inclusion of crystal field splittings for the valence orbitals can be treated similarly. We emphasize again that the computational cost is independent of such details, for comparable exciton radii.
	
	\section{1D Continuum approximation}
	\label{App:1D_Cont}
    
1D continuum approximation replaces the lattice dispersion of the electron and hole with parabolic bands, as if the lattice constant $a\rightarrow 0$. The corresponding counterpart of the 1D Hamiltonian with on-site Coulomb attraction becomes:
\begin{equation}
    H = -\frac{\hbar^2}{2m_e} \frac{\partial^2}{\partial x_e^2}-\frac{\hbar^2}{2m_h} \frac{\partial^2}{\partial x_h^2}- Ua \delta(x_e-x_h)
    \label{D1}
\end{equation}
where $x_e, x_h$ are the (continuous variable) positions of the electron and hole along the chain, and $m_e, m_h$ are their effective masses. Here \(a\) is a characteristic length scale, which we  identify with the lattice constant when comparing continuum and lattice predictions. 
    For our 1D lattice model, the conduction band dispersion is $E_c(k)\!=\Delta \!-2t_c\cos(ka)$, corresponding to an effective electron mass $m_e = {\hbar^2}/(2\,|t_c|\,a^2)$  in the limit $k=0$. For the valence band, we diagonalize the multi-orbital model to find the dispersion of the top valence band $E_v(k,+)$, and extract the effective hole mass from its $k\rightarrow 0$ limit, $ m_h = -\hbar^2/\frac{d^2 E_v(k,+)}{dk^2}|_{k=0}.$

Hamiltonian (\ref{D1}) can be factorized into center-of-mass (CM)  $X_{CM} = (m_e x_e + m_h x_h)/M$ where $M={m_e+m_h}$,  and relative $x=x_e - x_h$ components. CM motion with total momentum $K$ simply increases the total exciton energy by $\hbar^2 K^2/2M$. The $K=0$ exciton binding energy $E_{\rm bind}>0$ is obtained from :
	\begin{equation}
\left(-\frac{\hbar^2}{2\mu} \frac{\partial^2}{\partial x^2} - U a \delta(x)\right) \phi(x) = - E_{\rm bind} \phi(x),
	\end{equation}
	where $\mu = m_em_h/M$ is the reduced mass. This gives:
	\begin{equation}
		E_{\text{bind}} = \frac{\hbar^2 }{2\mu\xi^2} = \frac{\mu (U a)^2}{2 \hbar^2}.
	\end{equation}
    where the exciton radius is  \(\xi = \kappa^{-1} = \frac{\hbar^2}{\mu U a}\).

This can be generalized trivially to an indirect gap semiconductor: if the minimum gap corresponds to a momentum $K_g$, then the CM contribution to the exciton energy changes to $\hbar^2 (K-K_g)^2/2M$ while $E_{\text{bind}}$ is unaffected.

    	%\section{Detailed 2D Green's-Function Formalism}
    \section{2D continuum approximation}
    \label{App:2D_Cont}

Similar to the 1D case, the two–body continuum Hamiltonian separates into center–of–mass and relative parts, with the former contributing an additional $\hbar^2 ({\bm K}-{\bm K}_g)^2/2M$ to the exciton energy. To calculate the exciton binding energy, we need to regularize the on-site attraction into a circular well of radius $R_0$ whose integrated strength is $g$:
\begin{equation}
    V(r)= -\,\frac{g}{\pi R_0^2} \Theta(R_0 -r)
\end{equation}
where $\Theta$ is the Heaviside function. 
    We solve the radial Schr\"odinger equation to find $\phi(r) \sim J_0(\kappa r)$ if $r< R_0$, and $\phi(r) \sim K_0(\kappa r)$ if $r> R_0$, where $J_0, K_0$ are order-zero Bessel function of the first and  second kind, respectively, and $\kappa=\sqrt{{2\mu E_{\rm bind}}/{\hbar^2}}$. Matching the radial wavefunction and its derivative at $r=R_0$, we find:
    \begin{equation}
    E_{\rm bind}
    = \frac{\hbar^2}{2\mu R_0^2}\,
    \exp\!\Big[-\,\frac{4\pi\hbar^2}{\mu g}\Big].
    \label{E2}
    \end{equation}
    (Any additional pre-factor  of order 1 from the matching can be absorbed into the microscopic length $R_0$, which acts as a fitting parameter of the binding energy.)
    
    In the asymptotic limit $r\rightarrow \infty$, $K_0(\kappa r) \rightarrow {e^{-\kappa r}}/{\sqrt{r}}$ and we define the exciton radius as $\xi = 1/\kappa$. 
    
	\section{Continuum vs. lattice results in 2D}
    \label{App:exp_break}

\begin{figure}[t]
    \includegraphics[width=0.7\linewidth]{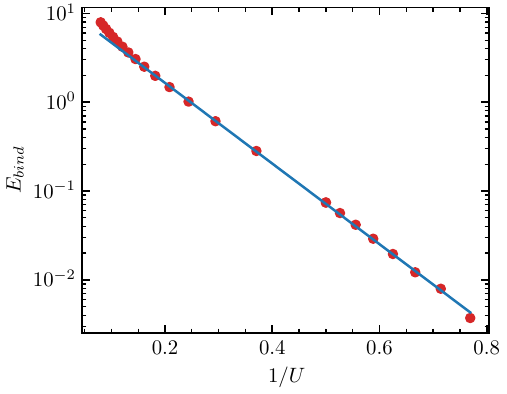}
    	\caption{Comparison between the lowest-exciton binding energy predicted by the 2D lattice model (red circles) and by the continuum model (blue line). Here, the parameters for the lattice model are $t_c = 1, t_{v1} = 0, t_{v2} = 0, t_{v3} = 1, t_{v4} = 0, \Delta = 7$, while those for the continuum model are obtained by fitting the lattice results with Eq. (\ref{E2}) for $U\in [0,2]$. The comparison is plotted on a semi-log scale vs.\ $1/U$. See text for more details.}
    	\label{fig:2D_1band_deviation_a}
    \end{figure}

\begin{figure}[b]
    \includegraphics[width=0.7\linewidth]{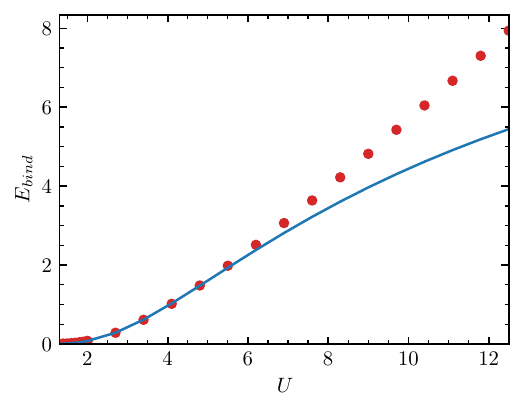} 
    	\caption{Same comparison as in Fig. \ref{fig:2D_1band_deviation_a} but on linear scale vs. $U$. As expected, the agreement is excellent for weak attraction $U$ but significant deviations appear for stronger attraction $U$.
        }
    	\label{fig:2D_1band_deviation_b}
    \end{figure}

Figure ~\ref{fig:2D_1band_deviation_a},~\ref{fig:2D_1band_deviation_b} show comparisons  between the lowest-energy exciton binding energy predicted by the 2D lattice model (red circles) and by its corresponding continuum model (blue solid line). The parameters for the lattice model are  $t_c = 1, t_{v1} = 0, t_{v2} = 0, t_{v3} = 1, t_{v4} = 0, \Delta = 7$, but qualitatively similar results are obtained for any other choice of parameters. For the continuum model, we use the lattice results for weak attraction  $U\in[0,2]$  to fit $R_0$ and $g$ from Eq. (\ref{E2}), and then generate the continuum prediction for all $U$.

 The upper panel shows the comparison on a semi-log scale against $1/U$, and verifies that for weak attraction $U$,  \(\ln E_{\text{bind}}\) is indeed linear in \(1/U\) as predicted by Eq. (\ref{E2}). However, the lower panel shows that on a linear scale, the two predictions start to disagree significantly for $U\gtrsim 6$ (for these parameters). This is consistent with the fact that for these lattice parameters, the bandwidth of the valence band is $W =9$. Indeed, disagreement with the continuum prediction is expected for  $|U|\gtrsim W$ when all the valence states (not just the ones at the top of valence band, which are described well by the continuum model) are involved in the exciton wavefunction.

    \section{Minimal variational approximation for the 2D strongly bound excitons}
    \label{App:1shell}

If the Coulomb attraction is so strong that the exciton is Frenkel-like, we expect a reasonable estimate to be obtained if we restrict the relative displacement between electron and hole to be $|{\bm \delta}|\le a$. One could then diagonalize the full Hamiltonian within this variational space, but the same results can be obtained efficiently from our method by setting all the propagators $\bm G_{\gamma, N}=0$ for $N\ge 2$, {\em i.e.} by choosing the cutoff at $M=2$.

In this case, the infinite recurrence relation for  the vectors $\bm G_{\gamma, N}$ derived in Section~\ref{sec:Conv_1D} simplifies to:
    \begin{align}
    (\underline\alpha_0+\mathcal U)\,\bm G_{\gamma, 0}
    &= \mathbf e_{\gamma}+\underline\beta_0(\mathbf K)\,\bm G_{\gamma, 1},\nonumber\\
    \underline\alpha_1(\mathbf K,z)\,\bm G_{\gamma, 1}
    &= \underline\gamma_1(\mathbf K)\,\bm G_{\gamma, 0},
    \label{eq:1shell_pair}
    \end{align}
so that $\bm G_{\gamma, 0}=\left[ \underline\alpha_0+\mathcal U - \underline\beta_0(\mathbf K) [\underline\alpha_1(\mathbf K,z)]^{-1}  \underline\gamma_1(\mathbf K)\right]^{-1} \mathbf e_{\gamma} $. The exciton energies are the poles of $\bm G_{\gamma, 0}$ located inside the gap: $z< E_{\rm gap}({\bm K})$, and are found from:
    \begin{equation}
    \det\!\Big[(z-\Delta + U)I_3
    -\;\underline\beta_0(\mathbf K)\;
    \underline\alpha_1(\mathbf K,z)^{-1}\;
    \underline\beta_0(\mathbf K)^{\dagger}
    \Big]=0.
    \label{eq:CF_pole_clean}
    \end{equation}
Here, we assumed the same attraction $U_\alpha =U$ for all orbitals, and used    $\underline\gamma_1(\mathbf K) =  \underline\beta_0(\mathbf K)^{\dagger} $ which follows because the Hamiltonian is Hermitian.

As discussed in the main text (see blue dashed line Fig.~\ref{fig:5}), this very simple approximation indeed provides results with decent quantitative agreement at sufficiently large $U$. Of course, if the exciton is spread over more sites, the agreement is improved by increasing $M$ accordingly. Clearly, the results must converge eventually if $M$ is sufficiently large -- this is the essence of the fully converged solution described in Section~\ref{sec:Conv_1D}.

\end{document}